\documentclass[%
 aip,
 jap,
 amsmath,amssymb,
reprint,%
floatfix,
]{revtex4-2}
\usepackage{setspace}
\usepackage[T1]{fontenc}
\usepackage[ngerman, english]{babel}
\usepackage{graphicx}
\usepackage{dcolumn}
\usepackage{bm}
\usepackage[hidelinks]{hyperref}
\hypersetup{colorlinks=false}

\usepackage[utf8]{inputenc}
\usepackage[T1]{fontenc}
\usepackage{mathptmx}
\usepackage{mathtools}
\usepackage{etoolbox}
\usepackage{float}
\usepackage{siunitx}
\usepackage{placeins}
\usepackage{capt-of}

\makeatletter
\gdef\@ptsize{1} 
\makeatother

\usepackage[ruled, ]{algorithm2e}

\usepackage{lipsum}
\usepackage{duckuments}
\usepackage{pgf}
\usepackage{import}
\usepackage{tikz}
\usepackage{nicefrac}
\usetikzlibrary{external}
\usetikzlibrary{shapes.arrows}
\usetikzlibrary{calc}
\usetikzlibrary{fit}
\DeclareMathOperator{\fft}{fft}
\DeclareMathOperator{\ifft}{ifft}
\DeclareMathOperator{\ifftshift}{ifftshift}
\DeclareMathOperator{\diag}{diag}
\DeclareMathOperator{\vecOP}{vec}

\usepackage{acro}

\DeclareAcronym{PSF}{
short = PSF ,
long = point spread function
}

\DeclareAcronym{PPT}{
short = PPT,
long = pulsed phase thermography
}

\DeclareAcronym{FMC}{
short = FMC ,
long = full matrix capture
}

\DeclareAcronym{FWHM}{
short = FWHM ,
long = full width at half maximum
}

\DeclareAcronym{ROI}{
short = ROI ,
long = region of interest,
long-plural-form = regions of interest
}

\DeclareAcronym{FFT}{
short = FFT ,
long = fast Fourier transform
}

\DeclareAcronym{PSR}{
short = photothermal SR,
long =  photothermal super resolution
}

\DeclareAcronym{SR}{
short = SR ,
long = super resolution
}

\DeclareAcronym{SNR}{
short = SNR ,
long = signal-to-noise ratio
}

\DeclareAcronym{ADMM}{
short = ADMM ,
long = alternating direction method of multipliers
}

\DeclareAcronym{OUT}{
short = OuT ,
long = object under test
}

\newcommand{\reffig}[1]{Fig.~(\ref{#1})}
\newcommand{\reffigl}[1]{Figure~(\ref{#1})}
\newcommand{\refeqq}[1]{Eq.~(\ref{#1})}
\newcommand{\refeqql}[1]{Equation~(\ref{#1})}

\newcommand{\refalgol}[1]{Algorithm~(\ref{#1})}

\newcommand{\plh}{{\,\mkern-2mu\times\mkern-2mu\,}}

\usepackage{array,tabularx,calc}

\newenvironment{conditions}[1][where:\hspace*{2mm}]
  {%
   #1\tabularx{\textwidth-\widthof{#1}}[t]{
     >{$}l<{$} @{${}\colon{}$} X@{}
   }%
  }
  {\endtabularx\\[\belowdisplayskip]}

\makeatletter
\def\@email#1#2{%
 \endgroup
 \patchcmd{\titleblock@produce}
  {\frontmatter@RRAPformat}
  {\frontmatter@RRAPformat{\produce@RRAP{*#1\href{mailto:#2}{#2}}}\frontmatter@RRAPformat}
  {}{}
}%
\makeatother
\begin{document}
\acuse{PSR}

\preprint{JAP22-AR-P1NDM2022-00743}

\title[Thermographic detection of internal defects using 2D photothermal super resolution reconstruction with sequential laser heating]{Thermographic detection of internal defects using 2D photothermal super resolution reconstruction with sequential laser heating}
\author{J. Lecompagnon}
\email{julien.lecompagnon@bam.de}
\homepage{https://www.bam.de/thermography}
\author{S. Ahmadi}%
\homepage{https://www.bam.de/thermography}
\author{P. Hirsch}%
\homepage{https://www.bam.de/thermography}
\affiliation{Bundesanstalt für Materialforschung und -prüfung (BAM), 12200 Berlin, Germany}%
\author{C. Rupprecht}
\affiliation{Technische Universität Berlin, 10623 Berlin, Germany}%

\author{M. Ziegler}
\homepage{https://www.bam.de/thermography}
\affiliation{Bundesanstalt für Materialforschung und -prüfung (BAM), 12200 Berlin, Germany}%

\date{\today}

\begin{abstract}
Thermographic photothermal super resolution reconstruction enables the resolution of internal defects/inhomogeneities below the classical limit which is governed by the diffusion properties of thermal wave propagation. Based on a combination of the application of special sampling strategies and a subsequent numerical optimization step in post-processing, thermographic super resolution has already proven to be superior to standard thermographic methods in the detection of one-dimensional defect/inhomogeneity structures. In our work, we report an extension of the capabilities of the method for efficient detection and resolution of defect cross sections with fully two-dimensional structured laser-based heating.
The reconstruction is carried out using one of two different algorithms which are proposed within this work. Both algorithms utilize the combination of several coherent measurements using convex optimization and exploit the sparse nature of defects/inhomogeneities as is typical for most nondestructive testing scenarios. Finally, the performance of each algorithm is rated on reconstruction quality and algorithmic complexity.
The presented experimental approach is based on repeated spatially structured heating by a high power laser. As a result, a two-dimensional sparse defect/inhomogeneity map can be obtained. In addition, the obtained results are compared with those of conventional thermographic inspection methods which make use of homogeneous illumination. Due to the sparse nature of the reconstructed defect/inhomogeneity map, this comparison is performed qualitatively.
\end{abstract}

\maketitle

\section{Introduction}

One of the main factors which governs the resolution limit of thermographic nondestructive testing methods is the diffusive nature of heat conduction in solids. This has an especially severe impact on resolving defect/inhomogeneity structures which lie deep below the surface of the \ac{OUT}. As a general rule, using well-established conventional thermographic testing methods, defects can only be fully resolved as long as their lateral extension is as large as they lie deep within the \ac{OUT}. With the application of photothermal \ac{SR} reconstruction techniques it has already been proven that this limit can be overcome.\\

\Acl{SR} imaging techniques have been successfully applied in several fields where there are resolution barriers to overcome. While geometrical \ac{SR} techniques enhance the spatial resolution of modern detectors to a sub-pixel accuracy \cite{Alam2000,Mandanici2019}, optical \acl{SR} techniques allow to overcome the classical Abbe diffraction limit  of optical imaging systems\cite{INGERMAN2018, Wang2019}.
However, currently established photothermal \ac{SR} reconstruction methods within the field of active thermographic materials testing are restricted to the reconstruction of one-dimensional defect structures\cite{Ahmadi2020, Ahmadi_2020, Burgholzer:17} or only approximate two-dimensional \ac{SR} by combining several one-dimensional illumination patterns\cite{Burgholzer2017}.\\

Furthermore, in this context it should be noted, that in contrast to the aforementioned methods relying on the spatially structured heating of the \ac{OUT}, there also exist reconstruction methods working with temporally structured illumination, which allow for a three-dimensional reconstruction of internal structures while also trying to eliminate thermal diffusion effects\cite{Kaiplavil2014,Tavakolian2017}.\\

Within this work we expand the experimental approach behind laser-based \ac{PSR} reconstruction towards the detection of two dimensional defects. This experimental approach is characterized by taking multiple independent measurements for a set of equidistantly distributed positions where the \ac{OUT} is heated using a single round laser spot with a high power laser. For each of those illuminations an independent measurement of the temperature response of the \ac{OUT} is recorded with an infrared camera.
In order to then reconstruct the internal defect/inhomogeneity pattern from the acquired set of independent measurements and gaining true two-dimensional information about the defect shape stripped from the blurring-effects of thermal diffusion, two different numerical methods for inverting the underlying severely ill-posed inverse problem are proposed. Subsequently, both methods are experimentally validated on a purpose-made additively manufactured sample with internal defect structures suitable to benchmark the resolution capabilities of each method. 

\section{Motivation on photothermal super resolution reconstruction}
The temporal progression of the front surface ($z=0$) temperature $T_{\text{meas}}$ of an \ac{OUT} exposed to a localized external heat flux $Q$ can be modeled as the sum of the initial temperature distribution on the \ac{OUT} front surface at $t=\SI{0}{\second}$ and the two-dimensional spatial convolution (denoted as $*_{x,y}$) of the thermal \ac{PSF} $\Phi_{\text{PSF}}$ and a heat source distribution $a$:
\begin{equation}
\label{eq:T0+psf}
    T_\text{meas}(x,y,z=0,t) = T_0(x,y) + \Phi_\text{PSF}(x,y,t) *_{x,y}\, a(x,y)\ .
\end{equation}
The thermal \ac{PSF} resembles the kernel function in the Green's-function like representation of the thermal diffusion differential equation. The \ac{PSF} can be calculated analytically for the special case of a defect-free thin plate with thickness $L$ as follows\cite{Cole2010, Ahmadi_2020}:
\begin{equation}
\begin{aligned}
   \label{eq:psf}
   \Phi_{\text{PSF}}(x,y,t) &= \left(\frac{2\,\mathrm{\hat{Q}}}{c_p\rho(4\mathrm{\pi}\alpha t)^{\nicefrac{n_\text{dim}}{2}}}\times \mathrm{e}^{-\frac{(x-\bar{x})^2+(y-\bar{y})^2}{4\alpha t}} \right.\\
    &\left.\times \sum_{n=-\infty}^{\infty} R^{2n+1}\mathrm{e}^{-\frac{(2nL)^2}{4\alpha t}} \right) *_t\, I_t(t),
\end{aligned}
\end{equation}
\vspace{-6mm}
\begin{table}[H]
\centering
\begin{conditions}
c_p 				& specific heat capacity \\
\rho 				& bulk density \\
\alpha 				& thermal diffusivity \\
n_\text{dim} 		& dimensionality of the heat flow: \newline
\begin{tabular}{p{3.8cm}p{1.5cm}}
\leftskip1em
point-wise excitation: 		& 	$n_\text{dim}\coloneqq 3$\\
\leftskip1em
line-wise excitation: 		&	$n_\text{dim}\coloneqq 2$\\
\leftskip1em
full-surface excitation: 	& 	$n_\text{dim}\coloneqq 1$\\
\end{tabular} \\
(\bar{x},\bar{y}) 	& coordinate centroid of the excitation area\\
R					& thermal wave reflection coefficient \newline 
(for typical metals $R\approx 1$) \\
*_t & denotes convolution in time \\
\mathrm{\hat{Q}} & amplitude of the external heat flux $Q$\\
I_t & temporal structure of $Q$. \\
\end{conditions}
\end{table}
\vspace{-5mm}
For the derivation of \refeqq{eq:psf} it is assumed, that the external heat flux $Q$ can be separated into its temporal structure given by $I_t$ and its spatial structure $ I_{x,y} *_{x,y}\, a_{\text{ext}}$ according to \refeqq{eq:Qdecomp}: 
\begin{equation}
    \label{eq:Qdecomp}
    Q(x,y,t) = \hat{Q} \times  I_{x,y}(x,y) *_{x,y}\, a_\text{ext}(x,y) *_t\, I_t(t) \ .
\end{equation}

Typically, $I_t$ with $I_t(t)\in \left[0,1\right]$ is a boxcar function with an active interval from $t\in[0,\,t_\text{pulse}]$, while $I_{x,y}(x,y)\in \left[0,1\right]$ encodes the intensity distribution of the external photothermal heating and $a_\text{ext}$ constitutes a distribution of Dirac delta impulses $\delta_i$ with a unit impulse at each excitation position:
\begin{equation}
a_\text{ext}(x,y)=\sum_i \delta_i(x_i,y_i) \ .
\end{equation}
While the temporal structure of the external heat flux is already considered within the \ac{PSF}, the spatial structure is incorporated into the heat source distribution $a$, which is defined as follows:
\begin{equation}
  \label{eq:a_structure}
  a(x,y) = I_{x,y}(x,y) *_{x,y}\, \left(a_\text{ext}(x,y)+a_\text{int}(x,y)\right)\ .
\end{equation}

\begin{figure}[htbp]
\includegraphics[scale=1]{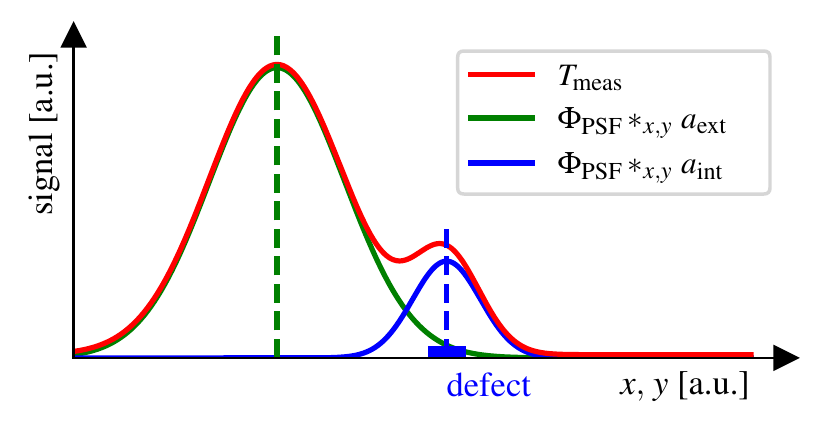}
\caption{Visualization of the components of the heat source distribution $a$: The measured temperature data $T_\text{meas}$ can be modeled according to \refeqq{eq:T0+psf} as the sum of the external heat source distribution $a_\text{ext}$ -- the external photothermal heating -- and the internal apparent heat source distribution $a_\text{int}$ -- the temperature deviations due to defects/inhomogeneities -- convolved with the thermal \ac{PSF} encoding the physics of heat conduction as well as the material parameters and geometry according to \refeqq{eq:T0+psf} and \refeqq{eq:a_structure}. $a_\text{ext}$ and $a_\text{int}$ are additionally shown as dashed lines.}
\label{fig:a_vis}
\end{figure}

Here, the additional term $a_\text{int}$ represents the spatial distribution of the internal \flqq apparent\frqq\ heat sources. In typical \ac{OUT}s, even though there are no active internal heat sources present, the fact that regions with internal defects/inhomogeneities show up as higher temperature regions in $T_\text{meas}$ lets them appear in the data as if they were independent heat sources (see \reffig{fig:a_vis}). This is due to the fact that they impede the local heat flow at their respective positions. It should be noted that there exist also scenarios in which defective regions show a lower temperature for which $a_\text{int}(x,y)$ is negative (apparent heat sinks). This is for example the case, if the \ac{OUT} includes inhomogeneities with a higher thermal effusivity than its bulk material. Both variants are equally covered by the presented modeling approach.

Therefore, if the heat source distribution $a$ can be reconstructed for the \ac{OUT}, then in turn  also the defect/inhomogeneity-structure of the \ac{OUT} is known, which in general makes the reconstruction of $a$ from \refeqq{eq:T0+psf} the overall goal of \ac{PSR} reconstruction. $a_\text{int}$ can be described as a Dirac delta pulse distribution as follows:
\begin{equation}
a_\text{int}(x,y)=\sum_i \zeta_i \times \delta(x_i,y_i)\ ,
\end{equation}
where $\zeta_i \in \left[0,1\right[$ serve as numerical weights in order to incorporate different defect signal strengths. These weights originate from the model assumption that all defects/inhomogeneities viewed as apparent heat sources feature the same \ac{PSF} as the external heating but depending on their cross section and depth attenuated by the factor $\zeta_i$. Since $\zeta_i$ can not easily be separated from $a_\text{int}$ within this modeling approach, by reconstructing the heat source distribution $a$ and therefore $a_\text{int}$ no information about the defect cross section in the depth plane can be gained. $a_\text{int}$ therefore only contains the in-plane information on the cross sections of the contained defects/inhomogeneities.\\

In order to solve the ill-posed problem as stated in \refeqql{eq:T0+psf} for the heat source distribution $a$, $n_m$ measurements with $m\in\{1, \dots, n_m\}$ are performed. For each measurement $m$ the photothermal heating pattern $a_\text{ext}$ is varied, leading to a system of $n_m$ equations to be solved simultaneously:
\begin{equation}
    \label{eq:core}
    \Phi_\text{PSF}(x,y,t) *_{x,y}\, a^m(x,y) =  T_\text{diff}^m(x,y,t)\ ,
\end{equation}
with $T_\text{diff}^m(x,y,t) = T^m_\text{meas}(x,y,t)-T_0^m(x,y)$.\\

\begin{figure*}[ht!]
\centering
\includegraphics[scale=1]{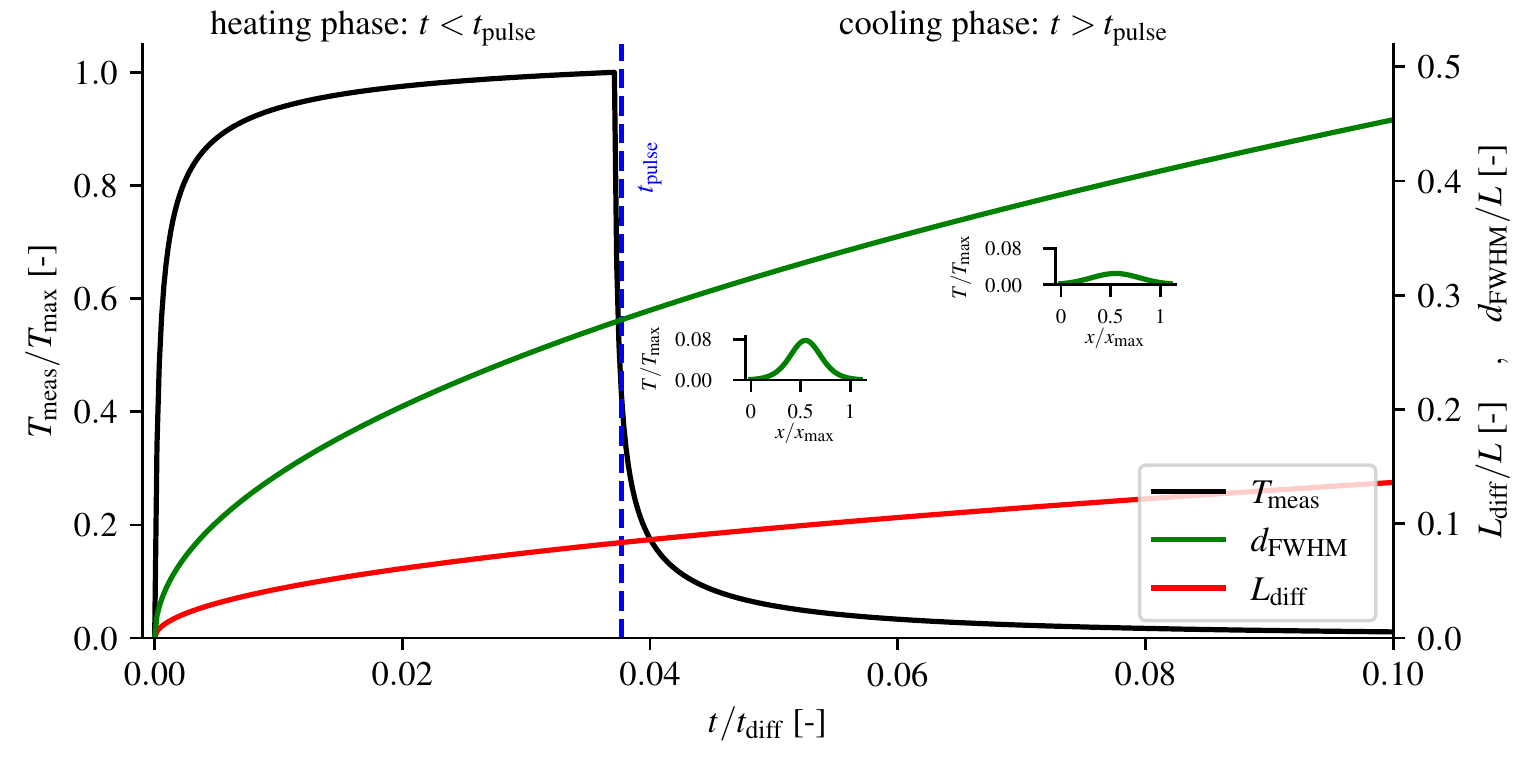}
\caption{Selection of $t_\text{eval}$: For eliminating the time dimension, the reconstruction is only performed for a single time step $t_\text{eval}$. This time step has to be chosen carefully with respect to the temperature signal strength (black line) for best \ac{SNR}, the thermal diffusion length (red line), and the homogeneity requirement in conjunction with the measurement density governed by $r_d$, which is influenced by the width $d_\text{FWHM}$ of the thermal \ac{PSF} $\Phi_\text{PSF}$ (green line). The inset figures show the spatial extent of $\Phi_\text{PSF}$ at $\nicefrac{t}{t_\text{diff}}=0.045$ and $0.065$ respectively. All shown graphs are normalized with the maximum temperature $T_\text{max}=\mathrm{max}(T_\text{meas})$ or the \ac{OUT} thickness $L$ and are plotted against the diffusion time $t_\text{diff}=\nicefrac{L^2}{\alpha}$ to exclude the dependency on material properties and geometry.}
\label{fig:homog}
\end{figure*}
Even though the external component $a_\text{ext}$ from \refeqq{eq:a_structure} is in principle known as prior information, it has shown to be beneficial to also simultaneously reconstruct the external component $a_\text{ext}$ besides $a_\text{int}$, leading to a blind reconstruction. If conversely a non-blind reconstruction approach is chosen, it is critical that the position and intensity profile of the external heating for each measurement is precisely known in order to not introduce reconstruction artifacts. In contrast, for a blind reconstruction this requirement does not apply, reducing the overall experimental complexity. However, in order to make it possible to separate $a_\text{ext}$ and $a_\text{int}$ without artifacts in a blind reconstruction context, it must hold true that the sum of the external excitation over all measurements is constant as demanded by \refeqq{eq:SRcond}:
\begin{equation}
	\label{eq:SRcond}
     I_{x,y}(x,y) *_{x,y}\, \sum_{m=1}^{n_m} a^m_\text{ext}(x,y) \approx const.
\end{equation}
This condition is based on the fact, that it can otherwise not be differentiated if a deviation from the mean is caused by an interference of a defect/inhomogeneity at that location or if it is caused by the non-uniformity of the sum of the external excitation.\\

Furthermore, to decrease the computational complexity it is advantageous to eliminate the time dependency of  \refeqq{eq:core} by only considering a single time step $t_\text{eval}$. When choosing a suitable time step, several factors have to be considered. In order to minimize measurement artifacts from the photothermal heating, a time step during the cooling phase $t_\text{eval}>t_\text{pulse}$ should be chosen. Due to the exponential nature of the cooling process, an early $t_\text{eval}$ leads to higher \ac{SNR}. Additionally the thermal diffusion length of the thermal wave $L_\text{diff}\propto t^{\nicefrac{1}{2}}$ increases with time with the thermal diffusivity $\alpha$ of the bulk material acting as the proportionality constant:
\begin{equation}
L_\text{diff}(t) = \sqrt{\alpha\times t} \ .
\end{equation}
Since \ac{PSR} is only able to gain in-plane information of the contained defects/inhomogeneities, only the planar projection of the defect/inhomogeneity distribution within the testing volume spanned by the \ac{ROI} and $L_\text{diff}$ can be reconstructed. Therefore, choosing $t_\text{eval}$ is a balance act between achieving sufficient \ac{SNR} and maximum detection depth for internal defects as is emphasized by \reffig{fig:homog}.
\section{Expanding to 2D-Reconstruction}
Currently well-established laser-based \ac{PSR} reconstruction methodology is restricted to reconstruction of defects along a single spatial dimension\cite{Ahmadi_2020, Ahmadi2021}. Here, the \ac{OUT} is heated using line-shape laser spots or patterns consisting of several laser lines either by scanning continuously over the \ac{ROI} or by step-wise scanning. Even though in this way only defects which vary only in a single dimension can be detected, the method has proven to highly increase the resolution limit of thermographic testing. One factor of its success is based on the fact, that one of the spatial dimensions can be collapsed by summation leading to a dramatic increase in \ac{SNR}, since all measured data in this collapsed dimension is redundant which greatly reduces the measurement noise, which is not possible for typical two-dimensional defects.\\

Other approaches on achieving two-dimensional \ac{SR} have been based on using movable slit-masks in conjunction with photothermal heating with the use of flash lamps. Here, the slit mask needs to not only be moved but also rotated to gain \ac{SR} in the second dimension and the information gain about the second spatial dimension is discretized to the different rotation angles of the slit mask\cite{Burgholzer2017}.
In order to expand the method to a two-dimensional reconstruction approach and to be sensitive in both spatial dimensions, a different heating strategy needs to be chosen.\\

\begin{figure}[ht!]
\centering
\includegraphics[scale=1]{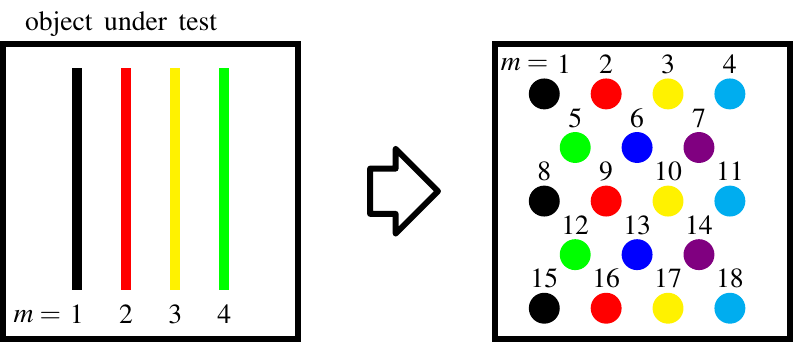}
\caption{Change in heating strategy to facilitate two-dimensional \ac{PSR} capabilities. In order to make achieving two-dimensional \ac{SR} possible, the heating pattern needs to be two-dimensional as well. Therefore, a switch from line-wise to a point-wise step scanning on a measurement grid of equilateral triangles is proposed.}
\label{fig:exp_trafo}
\end{figure}

If instead of scanning the \ac{OUT} with laser lines a sequential point-wise scanning approach is used, where the sample is iteratively scanned on an equally-spaced grid by a single laser spot, 2D-\ac{SR} can be achieved. In this scenario, $I_{x,y}$ can be modeled as a top-hat round laser spot with a diameter of $d_\text{spot}$. An equal spacing between all neighboring measurement positions can be achieved by arranging all measurements on a grid consisting of equilateral triangles of side length $r_d$ (c.f. \reffig{fig:exp_trafo}). A square \ac{ROI} with area $A_\mathrm{ROI,2D}$ and side length $L_\mathrm{ROI,1D}$ can therefore be covered by approximately $n_{m,2D} \approx\nicefrac{\sqrt{3}}{2}\times\nicefrac{A_\mathrm{ROI,2D}}{r_d^2}$ independent measurements:
\begin{align}
&& \frac{n_{m,2D}}{n_{m,1D}} &\approx \frac{\sqrt{3}\times A_\mathrm{ROI,2D}}{2\times L_\mathrm{ROI,1D}\times r_d} &&\\
\rightarrow && n_{m,2D} &\approx \frac{\sqrt{3}}{2}\times{n_{m,1D}}^2 \ . &&
\end{align}
In order to achieve sufficient uniformity according to \refeqq{eq:SRcond}, $r_d$ has to be chosen sufficiently small compared to the spatial \ac{FWHM} extension $d_\text{FWHM}$ of the \ac{PSF}, which has a direct effect on the amounts of independent measurements needed to cover the whole \ac{ROI}. If for a reconstruction along a single dimension ${n_{m,1D}=200}$ measurements are to be taken, for a 2D-reconstruction now ${n_{m,2D}\approx 34\,800}$ measurements are necessary to achieve a similar resolution for a fixed spacing of measurements $r_d$.\\

The \ac{FWHM} spot diameter $d_\text{FWHM}(t)$ for a \ac{PSF} with ${n_\text{dim}=2}$ is given by the following term considering the standard deviation $\sigma(t) = \sqrt{2\alpha\times t}$ of the \ac{PSF}:
\begin{equation}
d_\text{FWHM}(t) = 4 \sqrt{\ln(2)\alpha\times t} \ .
\end{equation}
Here, the time-dependency can be exploited such, that choosing a late-enough $t_\text{eval}$ will lead to a larger  $d_\text{FWHM}$ and therefore a larger necessary $r_d$, decreasing the amount of measurements to be performed to cover the \ac{ROI} and increase the viability of the method. However, this comes at the cost of lower \ac{SNR} and a reduced resolution limit. 
\section{Numerical Reconstruction}
The severely ill-posed fundamental equation of \ac{PSR} reconstruction as stated in \refeqql{eq:core} is formulated as a set of convolution problems $\Phi*_{x,y}\,a^m=T_\text{diff}^m$. Consequently, a suitable deconvolution algorithm needs to be applied to obtain a reconstruction of the internal heat source distribution $a_\text{rec}$ and therefore of the inner defect structure of the \ac{OUT}.
For a similar problem in matrix vector product form $\Phi\cdot a^m=T_\text{diff}^m$, this can be achieved by applying the iterative convex optimization \ac{ADMM} algorithm\cite{Boyd2010} in conjunction with $\ell_{2,1}$-norm and $\ell_2$-norm regularizing terms to constrain the solution space\cite{Haltmeier2016, Yu2013}:
\begin{equation}
\begin{aligned}
	\underset{a_\text{rec}}{\mathrm{minimize:}} \quad \nicefrac{1}{2}\ \left\Vert\Phi\right. &\left.\cdot \ a_\text{rec}^m - T_\text{diff}^m\right\Vert_2^2 \\
	&+ \lambda_{2,1}\left\Vert a_\text{rec}^m\right\Vert_{2,1} + \lambda_2\left\Vert a_\text{rec}^m\right\Vert_2^2\ , \end{aligned} 
\end{equation}
where $\left\Vert a_\text{rec}\right\Vert_{2,1}$ is the $\ell_{2,1}$-norm defined as:
\begin{equation}
	\left\Vert a_\text{rec}\right\Vert_{2,1} = \sum_m\sqrt{\sum_{x,y} |{a_\text{rec}^m}|^2} \ .
\end{equation}
$\lambda_2$ and $\lambda_{2,1}$ are numerical weights controlling the strength of the regularizing terms in comparison to the least squares minimization term $\nicefrac{1}{2}\left\Vert\Phi\cdot a_\text{rec}^m - T_\text{diff}^m\right\Vert_2^2$ and have to be empirically determined for optimal reconstruction results. Currently it is part of ongoing research, how to efficiently select the regularizer weights in the context of \ac{PSR} reconstruction using machine learning techniques\cite{Ahmadi2020c, Ahmadi2021a}.\\

While the $\ell_2$-norm constrains the amplitude of the reconstruction result, the $\ell_{2,1}$-norm incentivizes the reconstruction of (joint-)sparse defect/inhomogeneity structures, which effect multiple measurements. Since the defect/inhomogeneity structure of the \ac{OUT} does not change for all measurements and defects/inhomogeneities are typically scarcely distributed within the \ac{ROI}, this regularization approach greatly increases the reconstruction quality.\\
Additionally the \ac{ADMM} algorithm introduces a penalty parameter $\rho$ (see detailed \ac{ADMM}-implementations in \refalgol{alg:SMS} and \refalgol{alg:FFT}), which balances the effect of the least squares minimization term against the regularizing terms. This parameter can be determined programmatically using the L-curve method\cite{Hansen1994}.\\

Finally, the internal heat source distribution $a_\text{int}$ widened by $I_{x,y}$ can be extracted by summing over all measurements where the external component $a_\text{ext}$ is eliminated due to the homogeneity constraint as stated in \refeqq{eq:SRcond}:
\begin{equation}
\label{eq:sum_nm}
n_m \times I_{x,y}(x,y)*_{x,y}\, a_\text{int}(x,y) + const. = \sum_{m=1}^{n_m} a_\text{rec}^m(x,y) \ .
\end{equation}
Furthermore, the constant term in \refeqq{eq:sum_nm} originating from \refeqq{eq:SRcond} vanishes due the application of the $\ell_2$-norm within the \ac{ADMM} algorithm as it will converge to zero over the performed $n_\text{iter}$ iterations.

Within the following sections, two methods are proposed for transforming the spatial convolution problem into a multiplicative form, such that the aforementioned inversion via the \ac{ADMM} algorithm can be performed\cite{Lecompagnon2021,Lecompagnon2021a}.

\subsection{Sparse Matrix Stacking}
In order to transform the fundamental equation from a convolution problem into a multiplicative form to be able to use the \ac{ADMM} algorithm for inversion, first the spatial dimensions have to be combined into one single dimension $r\in \{1, \dots, n_x\cdot n_y\}$ by vectorizing:
\begin{align}
	\vecOP(\Phi \in \mathbb{R}^{n_x \times n_y}) 				&= \Phi_r \in \mathbb{R}^{n_x\cdot n_y \times 1}\\
	\vecOP(a^m \in \mathbb{R}^{n_x \times n_y}) 				&= a_r^m  \in \mathbb{R}^{n_x\cdot n_y \times 1}\\
	\vecOP(T_\text{diff}^m \in \mathbb{R}^{n_x \times n_y}) 	&= T_r^m  \in \mathbb{R}^{n_x\cdot n_y \times 1} \ .
\end{align}

After that, a convolution matrix operator $h(\cdot)$ can be introduced. This operator constructs for an input matrix of shape $\Phi_r$ a convolution matrix $h(\Phi_r) \in \mathbb{R}^{2n_x\cdot n_y-1\,\times\, n_x\cdot n_y}$, such that:
\begin{equation}
\Phi_r *_{r}\, a_r^m = T_\text{diff}^m \quad\Leftrightarrow\quad h(\Phi_r)\cdot a_r^m = \begin{bmatrix} \underline{0}\\ T_r^m \\ \underline{0}  \end{bmatrix} = T_{r0}^m \ .
\end{equation}
Here, $T_{r0}^m \in \mathbb{R}^{2n_x\cdot n_y-1}$ comprises the measured surface temperature data $T_r^m$ symmetrically padded with $\nicefrac{n_x\cdot n_y}{2}$ zeros. The convolution matrix $h(\Phi_r)$ constitutes a lower triangular matrix with Toeplitz-structure and is therefore quite sparsely populated.
Since all $n_m$ measurements need to be considered simultaneously for the reconstruction, the $n_m$ equations can be stacked on top of each other as follows:
\begin{equation}
	\label{eq:stackedEq}
	\setlength{\arraycolsep}{0pt}
	\setlength{\delimitershortfall}{0pt}
	H\cdot A = \begin{bmatrix}\,\fbox{$h$} &  & \underline{0}\, \\ \, & \fbox{$h$} & \, \\ \,\underline{0} &  & \fbox{$h$}\, \\ \end{bmatrix}\cdot\begin{bmatrix} a_r^1\\\vdots\\a_r^{n_m }\end{bmatrix} = \begin{bmatrix} T^1_{r0}\\\vdots\\T^{n_m }_{r0}\end{bmatrix} = T_{R0} \ ,
\end{equation}
with $h = h(\Phi_r)$, $H \in \mathbb{R}^{(2n_x\cdot n_y-1)\cdot n_m\,\times\, n_x\cdot n_y\cdot n_m}$, $A \in \mathbb{R}^{n_x\cdot n_y\cdot n_m}$ and $T_{R0} \in \mathbb{R}^{(2n_x\cdot n_y-1)\cdot n_m}$. Even though the dimensional size of the matrix $H$ is orders of magnitudes larger than the original data size of $n_x\cdot n_y\cdot n_m$ data points, it is as a diagonal matrix of sparse component matrices very sparse itself (only $\approx\nicefrac{1}{2n_m}$ of the entries in $H$ contain a non-zero value), making it possible to be still efficiently stored and handled on modern computer hardware.\\

In summary, the following minimization problem emerged:
\begin{equation}
\label{eq:minA}
	\underset{A}{\mathrm{minimize:}} \quad \nicefrac{1}{2}\ \left\Vert H\cdot A-T_{R0}\right\Vert_2^2 + \lambda_{2,1} \left\Vert A\right\Vert_{2,1} + \lambda_2 \left\Vert A\right\Vert^2_2 \ ,
\end{equation}
for which the algorithm presented as \refalgol{alg:SMS} can be applied to solve for $A$ and ultimately reconstruct the internal defect/inhomogeneity map as encoded in $a_\text{int}$.\\
\begin{figure}[!htbp]
  \centering
        \begin{algorithm}[H]
        \SetKwInOut{Input}{input}\SetKwInOut{Output}{output}
        \Input{$H$, $T_{R0}$, $\rho$, $\lambda_{2,1}$, $\lambda_{2}$, $n_\text{iter}$}
        \Output{$a_{\text{rec}}$}
        \begin{onehalfspacing}
        \SetKwFunction{FMain}{$prox_{\ell_{21}+\ell_{2}}$}
        \SetKwProg{Fn}{function}{:}{}
        \Fn{\FMain{$l$, $\lambda_{2,1}$, $\lambda_{2}$}}{
            $p[r, m] \leftarrow max(0 , 1- \frac{\lambda_{2,1}}{\sqrt{\sum_{m=1}^{n_m }|l[r, m]|^2}})\frac{l[r, m]}{1+\lambda_{2}}$ \\
        \KwRet $p$}
        \vspace{2mm}
        $x^{(0)} \leftarrow \text{random\_uniform}_{[0,1[}\in \mathbb{R}^{n_x\cdot n_y\cdot n_m \times1}$\\
        $z^{(0)} \leftarrow \text{random\_uniform}_{[0,1[}\in \mathbb{R}^{n_x\cdot n_y\cdot n_m \times1}$\\
        $u^{(0)} \leftarrow x^{(0)}-z^{(0)}$\\
        \For{$\mathrm{k}\leftarrow 1$ \KwTo $\mathrm{n_\text{iter}}$}{
            $x^{(k)} \leftarrow (H^TH+\rho I)^{-1}(H^TT_{R0}+\rho(A^{(k)}-u^{(k)}))$\\
            $l^{(k)} \leftarrow$ reshape\ $x^{(k)}+u^{(k-1)}$ \KwTo $\mathbb{R}^{n_x\cdot n_y \times n_m }$ \\
            $p^{(k)} \leftarrow prox_{\ell_{21}+\ell_{2}}(l^{(k)},\, \lambda_{2,1}/\rho,\, \lambda_{2}/\rho)$\\
            $z^{(k)} \leftarrow$ reshape $p^{(k)}$ \KwTo $\mathbb{R}^{n_x\cdot n_y\cdot n_m \times1}$\\
            $u^{(k)} \leftarrow u^{(k-1)}-A^{(k)}$\\
        }
     	$a_{\text{rec}} \leftarrow$ $\sum_{m=1}^{n_m}\left( \text{reshape\ } A^{(n_\text{iter})}\ \KwTo\ \mathbb{R}^{n_x\times n_y\times n_m}\right)$\\
        \KwRet $a_{\text{rec}}$\\
        \end{onehalfspacing}
        \label{alg:SMS}
        \caption{Sparse Matrix Stacking Reconstruction}
        \end{algorithm}
\end{figure}

The computational complexity of \refalgol{alg:SMS} is mainly dominated by repeated matrix vector multiplications. Na\"ively this will run in at most $\mathcal{O}((2n_x\cdot n_y-1)^2\cdot n_x\cdot n_y\cdot n_m^3)$ time, but due to the high degree of sparsity involved, this establishes only a very conservative upper bound\cite{Yuster2005}.
\subsection{Minimization in the Frequency Domain}
Another possible way to transform a spatial convolution problem into a multiplicative form is to switch into the spatial frequency domain. This can be achieved by means of a two-dimensional \ac{FFT} as follows:
\begin{equation}
\Phi_{\text{PSF}}\ *_{x,y}\, a^m \ \Leftrightarrow \ \ifft\Big(\overline{\fft\big(\ifftshift\big(\Phi_{\text{PSF}}\big)\,\big)}\odot\, \fft\big(a^m\big) \Big) \ ,
\end{equation}
where $\fft$ indicates the two-dimensional \ac{FFT} in $x$ and $y$, $\ifft$ its inverse, $\ifftshift$ a function which swaps the first quadrant with third and the second with the fourth quadrant of the input matrix, overlines designate taking the complex conjugate and $\odot$ indicates element-wise multiplication (Hadamard product). In order to get rid of the element-wise multiplication and transform the problem to a matrix vector multiplication problem, the following substitution can be applied:
\begin{equation}
\begin{aligned}
&\ifft\Big(\overline{\fft\big(\ifftshift\left(\Phi_{\text{PSF}}\right)\,\big)}\odot\, \fft\big(a\big) \Big) \\
&\Leftrightarrow \quad \diag\bigg(\vecOP\Big(\overline{\fft\big(\text{ifftshift}\left(\Phi_{\text{PSF}}\right)\big)}\Big)\bigg) \cdot \vecOP\big(\fft\left(a\right)\big) \ .
\end{aligned}
\end{equation}
Since the constraints governing the selection of the regularization terms for the problem inversion, like the sparse distribution of defects/inhomogeneities, are only applicable in the spatial domain, the problem has to be transformed back from the spatial frequency domain in order to apply their effect. This leads to one invocation of the $\fft$ and one of the $\ifft$ operation necessary per iteration.
\vspace*{-0.2cm}
\begin{figure}[H]
  \centering
        \begin{algorithm}[H]
        \SetKwInOut{Input}{input}\SetKwInOut{Output}{output}
        \Input{\ $\Phi_{\text{PSF}}$, $T_{\text{diff}}$, $\rho$, $\lambda_{2,1}$, $\lambda_{2}$, $n_\text{iter}$}
        \Output{\ $a_\text{rec}$}
        \begin{onehalfspacing}
        \SetKwFunction{FMain}{$prox_{\ell_{21}+\ell_{2}}$}
        \SetKwProg{Fn}{function}{:}{}
        \Fn{\FMain{$l$, $\lambda_{2,1}$, $\lambda_{2}$}}{
            $p[i, j, m] \leftarrow max(0 , 1- \frac{\lambda_{2,1}}{\sqrt{\sum_{m=1}^{n_m }|l[i, j, m]|^2}})\frac{l[i, j, m]}{1+\lambda_{2}}$ \\
        \KwRet $p$}
        \vspace{2mm}$A \leftarrow \diag\Big($reshape\ $\overline{\fft\left(\text{ifftshift}( \Phi_\text{PSF}\right)} $ \KwTo $\mathbb{R}^{n_x\cdot n_y\cdot n_m \times1}\Big)$ \\
        $b \leftarrow$ reshape\ $\fft(T_{\text{diff}})$ \KwTo $\mathbb{R}^{n_x\cdot n_y\cdot n_m \times1}$ \\
        $x^{(0)} \leftarrow \text{random\_uniform}_{[0,1[}\in \mathbb{R}^{n_x\cdot n_y\cdot n_m \times1}$\\
        $z^{(0)} \leftarrow \text{random\_uniform}_{[0,1[}\in \mathbb{R}^{n_x\cdot n_y\cdot n_m \times1}$\\
        $u^{(0)} \leftarrow x^{(0)}-z^{(0)}$\\
        \For{$\mathrm{k}\leftarrow 1$ \KwTo $n_\text{iter}$}{
        
            $x^{(k)} \leftarrow (A^TA+\rho I)^{-1}(A^Tb+\rho(z-u))$\\
            $l^{(k)} \leftarrow$ reshape\ $ \ifft(x^{(k)}+u^{(k-1)})$ \KwTo $\mathbb{R}^{n_x \times n_y \times n_m }$ \\
            $p^{(k)} \leftarrow prox_{\ell_{21}+\ell_{2}}(l^{(k)},\, \nicefrac{\lambda_{2,1}}{\rho},\, \nicefrac{\lambda_{2}}{\rho})$\\
            $z^{(k)} \leftarrow$ reshape\ $\fft(p^{(k)})$ \KwTo $\mathbb{R}^{n_x\cdot n_y\cdot n_m \times1}$\\
            $u^{(k)} \leftarrow u^{(k-1)}-z^{(k)}$\\
        }
        
        $a_\text{rec} \leftarrow \sum_{m=1}^{n_m }\left( \text{reshape\ } \ifft(z^{(n_\text{iter})})\ \KwTo\ \mathbb{R}^{n_x\times n_y \times n_m }\right)$\\
        \KwRet $a_\text{rec}$\\
        \end{onehalfspacing}
        \label{alg:FFT}
        \caption{Frequency Domain Reconstruction}
        \end{algorithm}
\end{figure}
\vspace*{-0.2cm}
The computational complexity of \refalgol{alg:FFT} is mainly dominated by several invocations of the \ac{FFT} per iteration. It will run in $\mathcal{O}( n_x\cdot n_y\cdot n_m\cdot \log(n_x\cdot n_y\cdot n_m))$ time, which is significantly less than the sparse matrix stacking minimization approach presented before for most circumstances.
\section{Experimental Setup}
In order to test the experimental approach in conjunction with both proposed inversion algorithms as an extension to laser-based \ac{PSR} reconstruction, a purpose-made sample additively manufactured from 316L stainless steel has been examined. Within the \ac{ROI} as shown in \reffig{fig:exp_out} lie multiple internal defects in the form of cubical voids with side length $\SI{2}{\milli\metre}$, filled with unfused metal powder from the manufacturing process. The square \ac{OUT} features a side length of $\SI{58.5}{\milli\metre}$, a thickness of $L=\SI{4.5}{\milli\metre}$, while its fused bulk material comes with a thermal conductivity of $k=\SI{15}{\watt\per\metre\kelvin}$, a density of $\rho=\SI{7950}{\kilo\gram\per\metre\cubed}$ and a specific heat capacity of $c_p=\SI{502}{\joule\per\kilo\gram\per\kelvin}$  and therefore an inferred thermal diffusivity of $\alpha=\SI{3.76e-6}{\metre\squared\per\second}$.\, \cite{316lDatasheet,ASMInternational2009}\\

The \ac{ROI} is chosen such, that it covers several pairs of defects with a spacing of $\SI{0.5}{\milli\metre}$, $\SI{1}{\milli\metre}$, $\SI{2}{\milli\metre}$ and $\SI{4}{\milli\metre}$ between them. This allows not only for testing if all defects are detected by the algorithm, but also if all defects can be detected as individual defects resulting in a measure of the resolution capabilities of the method.
\begin{figure}[H]
\centering
\includegraphics[scale=0.9]{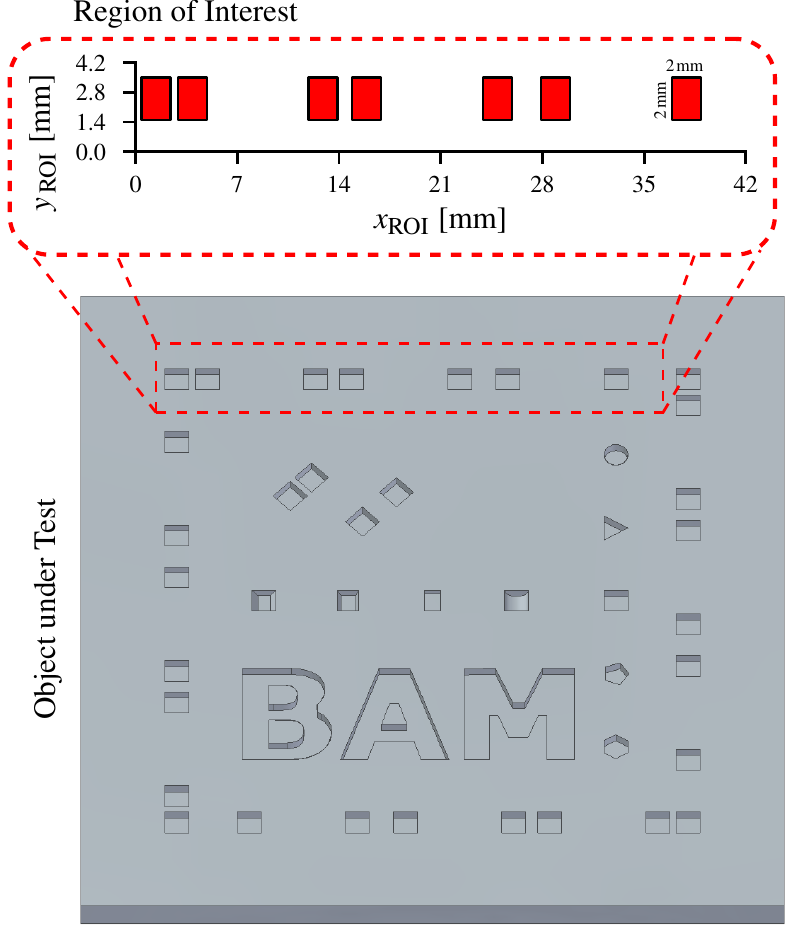}
\caption{Section view at a depth of $z=\SI{1.5}{\milli\metre}$ of the \ac{OUT}: The \ac{OUT} is additively manufactured from 316L stainless steel. The \ac{ROI} for the experiment is marked as a dashed red box. A detailed map of the \ac{ROI}  with the locations of the defective regions shown as red boxes is overlayed on top. The cubical voids within the \ac{ROI} have a side length of $\SI{2}{\milli\metre}$ and are fully covered by a defect-free $\SI{0.5}{\milli\metre}$ thick cover layer.}
\label{fig:exp_out}
\end{figure}
As a photothermal heating source, a $\hat{Q}_\text{max}=\SI{500}{\watt}$ fiber-couple diode laser with a wavelength of $\lambda=\SI{940}{\nano\metre}$ is used. With this the \ac{OUT} front surface can be illuminated via reflecting off a dichroitic mirror (see \reffig{fig:exp_setup}), which is highly reflective for the laser wavelength but transparent for midwave infrared radiation. The surface temperature is subsequently measured with a midwave infrared camera with a detector size of $n_\text{pix}= 1280\plh\SI{1024}{pixel}$ at a spatial resolution of $\Delta x, \Delta y=\SI{52}{\micro\meter}$ per pixel with a framerate of $f_\text{cam}=\SI{100}{\hertz}$. Two linear translational stages are utilized to move the sample with respect to the laser and allow for scanning of the whole surface of the \ac{OUT}.
\begin{figure*}
\centering
\includegraphics[width=16.9cm]{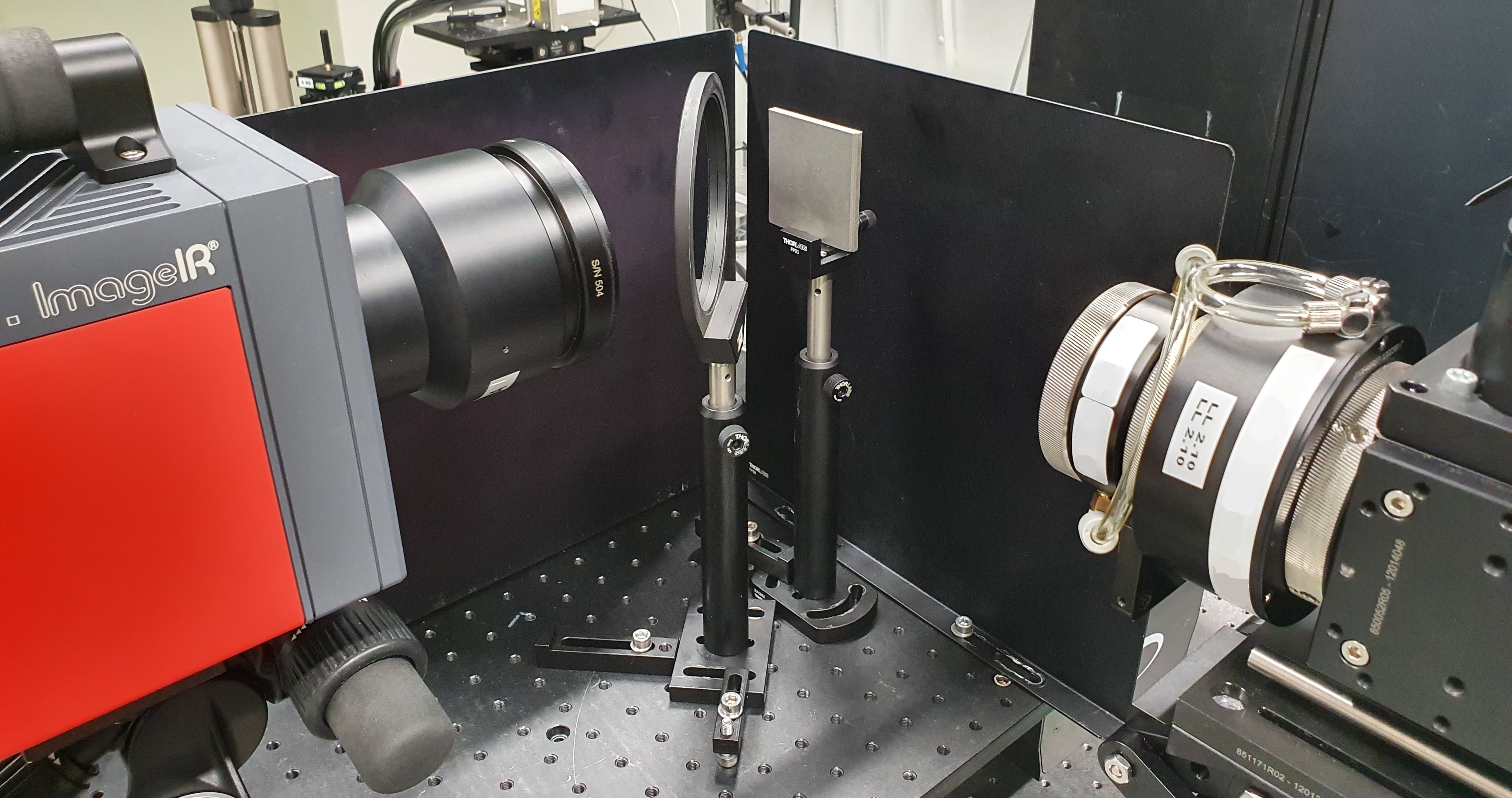}
\caption{Experimental setup: The \ac{OUT} (middle) is heated by a fiber coupled diode laser with interchangeable projection objectives (right) and a maximum optical output power of $\hat{Q}_\text{max}=\SI{500}{\watt}$ via a dichroitic mirror (middle). This mirror is highly reflective for the laser wavelength and transparent for midwave infrared radiation. The surface temperature is recorded with a midwave infrared camera with a detector size of $n_\text{pix}= 1280\times\SI{1024}{pixel}$ at a spatial resolution of $\Delta x, \Delta y=\SI{52}{\micro\meter}$ per pixel and a framerate of $f_\text{cam}=\SI{100}{\hertz}$ (left). Scanning is achieved by utilizing two translational stages which move the \ac{OUT} within the image plane of the laser focus objective.}
\label{fig:exp_setup}
\end{figure*}
\section{Experimental Results}
%
%
For a regular measurement grid of equilateral triangles of side length $r_d=\SI{0.743}{\milli\metre}$ across the \ac{ROI} with $n_m=403$ vertices/independent measurement positions arranged in $7$ rows with $54$ or $53$ positions each and subsequent reconstruction using the sparse matrix stacking approach, the resulting reconstruction of $a_\text{rec}$ is shown in \reffig{fig:res_sms}. At each grid point a single measurement has been recorded after an illumination with a laser spot with spot size $d_\text{spot}=\SI{0.6}{\milli\metre}$ at $\hat{Q}=\SI{15}{\watt}$. To eliminate the time dimension, an evaluation time of $t_\text{eval}=\SI{500}{\milli\second}$ has been taken into account. This corresponds to a thermal diffusion length of $L_\text{diff}\approx\SI{1.5}{\milli\metre}$.\\

The reconstruction result for applying the sparse matrix stacking algorithm is shown in \reffig{fig:res_sms}. For this reconstruction the \ac{ADMM}-parameters $\lambda_{2,1}=1570$, $\lambda_2=100$, $\rho=16$ and $n_\text{iter}=400$ have been used. As can be seen, all defects have been resolved as independent defects and the measurement noise has been successfully suppressed. The reconstruction of the exact defect geometry is still to be improved. For reference, in \reffig{fig:res_flash} the measured data for illuminating the whole \ac{OUT} front surface homogeneously using the same setup as for the point-wise illumination with $\hat{Q}=\SI{500}{\watt}$ and a pulse length of $t_\text{pulse}=\SI{500} {\milli\second}$ is presented.\\

\begin{figure}[H]
\centering
\includegraphics[scale=1]{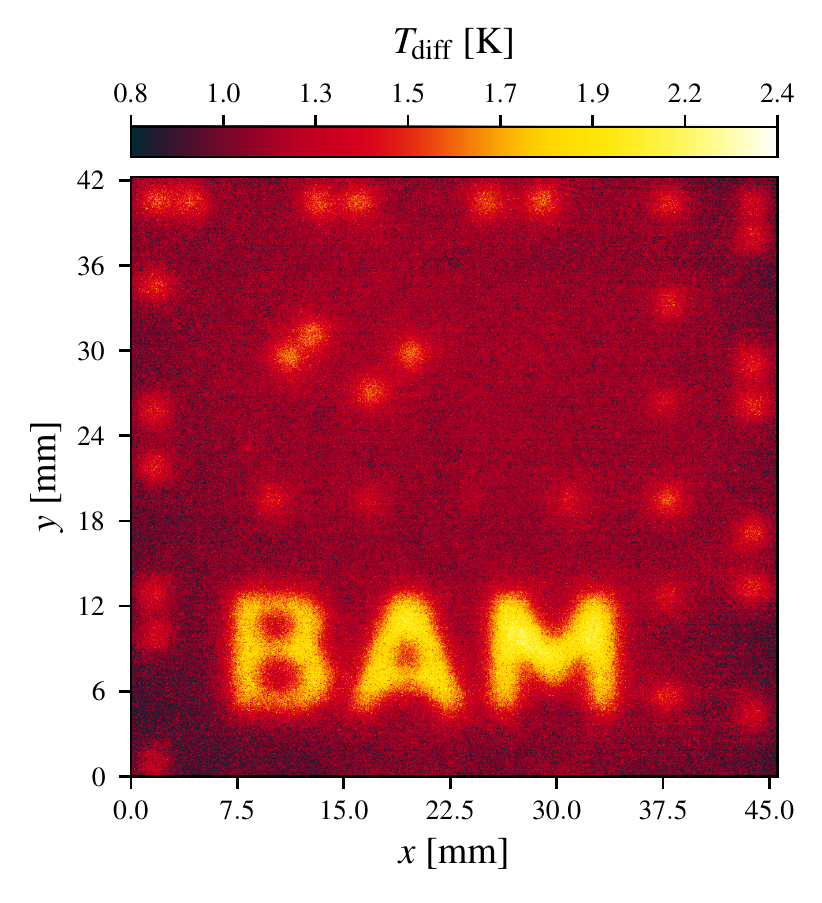}
\caption{Reference measurement for homogeneous illumination of the whole \ac{OUT} surface with a laser pulse with $\hat{Q}=\SI{500}{\watt}$ and a pulse length of $t_\text{pulse}=\SI{500}{\milli\second}$ evaluated at $t_\text{eval}=\SI{500}{\milli\second}$. For generation of this data the same setup as shown in \reffig{fig:exp_setup} has been used.}
\label{fig:res_flash}
\end{figure}

\begin{widetext}
\begin{center}
\includegraphics[scale=1]{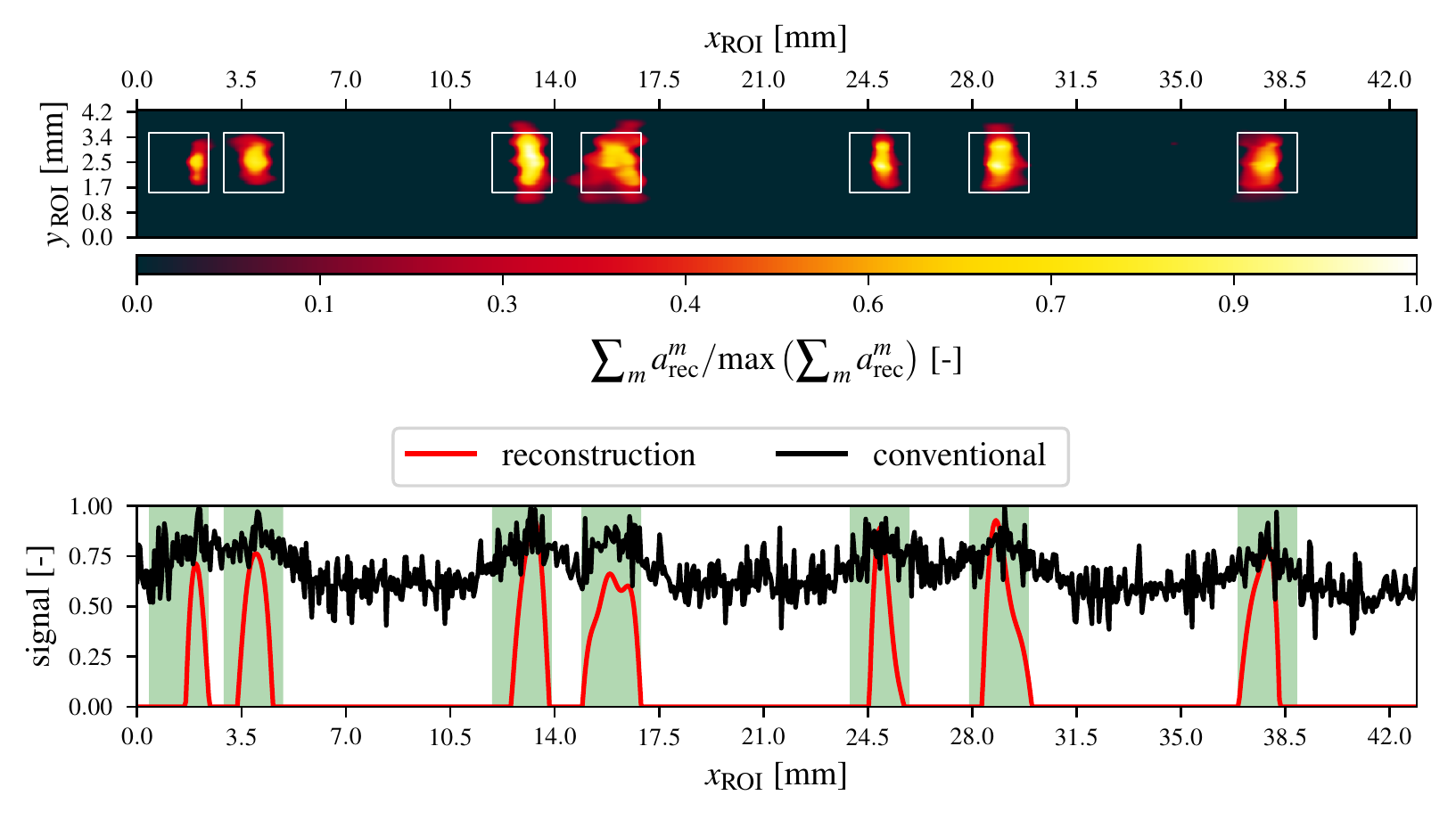}
\captionof{figure}{Reconstruction result for $a_\text{rec}$ at $t_\text{eval}=\SI{500}{\milli\second}$ reconstructing using the sparse matrix stacking method. For this reconstruction the \ac{ADMM}-parameters ${\lambda_{2,1}=1570}$, ${\lambda_2=100}$, ${\rho=16}$ and $n_\text{iter}=400$ have been used. The top figure shows $a_\text{rec}$ with white boxes overlayed at the locations where the defective regions are. The bottom figure shows a sectional view of the data for ${y_\text{ROI}=\SI{2.5}{\milli\metre}}$. The light green shaded areas mark the defective region. For reference, the normalized conventionally acquired data at the same coordinates from \reffig{fig:res_flash} is plotted as well.}
\label{fig:res_sms}

\includegraphics[scale=1]{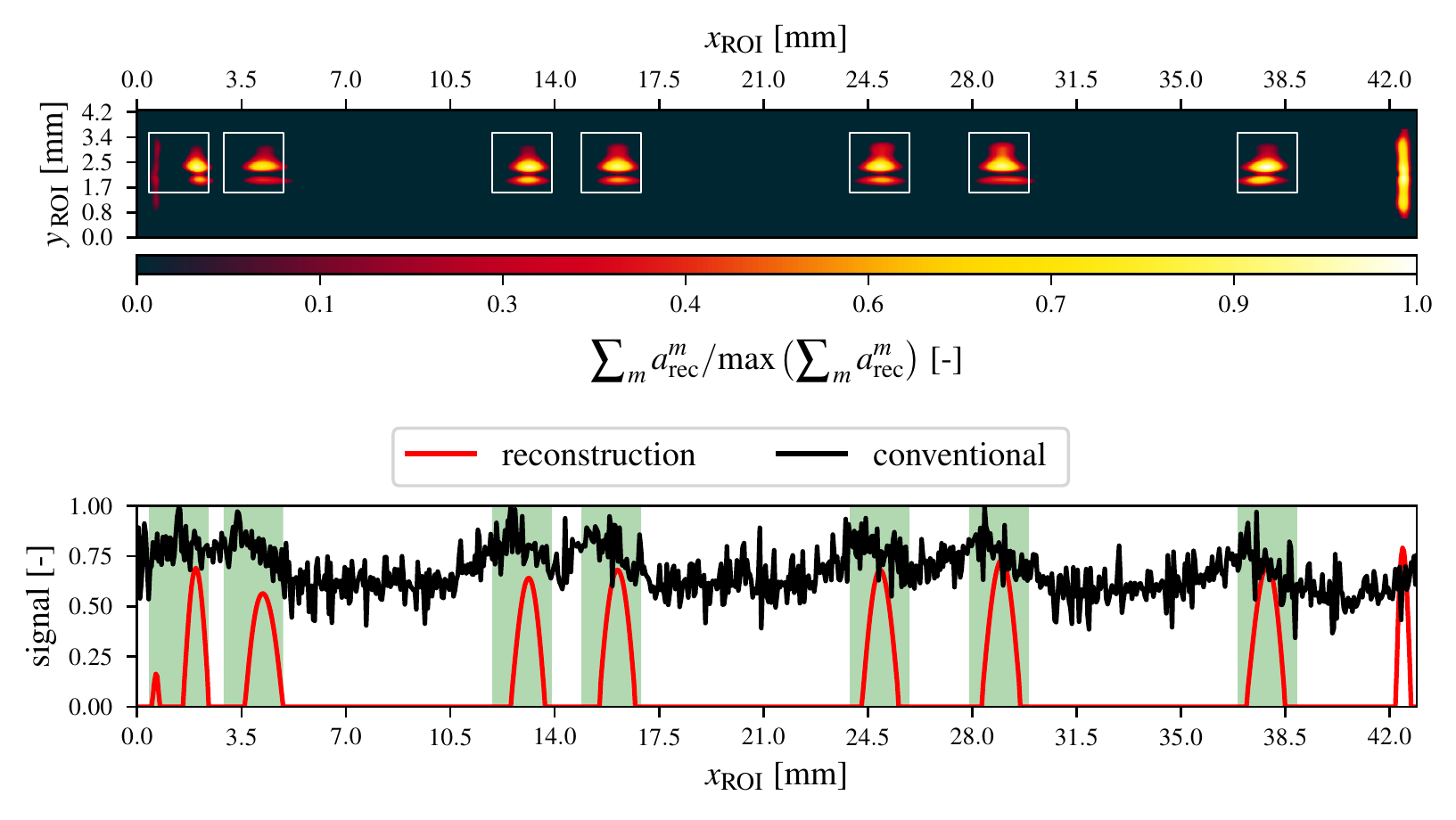}
\captionof{figure}{Reconstruction result for $a_\text{rec}$ at $t_\text{eval}=\SI{700}{\milli\second}$ reconstructing within the frequency domain. For this reconstruction the \ac{ADMM}-parameters ${\lambda_{2,1}=27}$, ${\lambda_2=500}$, ${\rho=16}$ and ${n_\text{iter}=400}$ have been used. The top figure shows $a_\text{rec}$ with white boxes overlayed at the locations  where the defective regions are. The bottom figure shows a sectional view of the data for ${y_\text{ROI}=\SI{1.92}{\milli\metre}}$. The light green shaded areas mark the defective region. For reference, the normalized conventionally acquired data at the same coordinates from \reffig{fig:res_flash} is plotted as well.}
\label{fig:res_fft}
\end{center}
\end{widetext}

The reconstruction result for reconstructing $a_\text{rec}$ within the spatial frequency domain is shown in \reffig{fig:res_fft}. Here, the \ac{ADMM}-parameters $\lambda_{2,1}=27$, $\lambda_2=500$, $\rho=16$ and $n_\text{iter}=400$ have been used to achieve the shown reconstruction results. With reconstruction within the spatial frequency domain, a speedup of approximately $50$ times has been found for the reconstruction within the given \ac{ROI}, while this method is also much more sensitive to perturbation, making it harder to find suitable set of parameters for the reconstruction.\\

In contrast to the sparse matrix stacking reconstruction result, this method also reconstructs the edges of the illuminated area as defects as can be seen near the left and right borders of the \ac{ROI}. This can be explained by the local violation of the necessary condition for \ac{SR} reconstruction given in \refeqq{eq:SRcond}. The sparse matrix stacking inversion method has been experienced to show much less dependence on the smoothness of the external excitation pattern. Even though both numerical inversion methods mathematically approximate the true solution to the same reconstruction problem, both methods show considerably different results. This can not only be explained due to performing the inversion either in the flattened spatial domain (sparse matrix stacking) or the spatial frequency domain (\ac{FFT}-based method) but also by the high non-linearity and filtering properties of the used solver in form of the \ac{ADMM} algorithm and the applied regularizing terms. How to exploit those differences to improve the reconstruction quality is still part of ongoing research.\\

In order to better assess the resolution gain of the proposed \ac{PSR} reconstruction methods compared to well-established conventional methods, a qualitative comparison of different methods can be found in \reffigl{fig:res_comp}. Here, the reconstruction results of both proposed \ac{PSR} reconstruction methods is shown next to raw data acquired after homogeneous illumination (c.f. \reffigl{fig:res_flash}) as-is and after different post-processing steps. The homogeneous illumination data is recorded within the same setup as shown in \reffigl{fig:exp_setup}. The additional post-processing steps shown contain a difference thermogram where from the measured raw data $T_\text{diff}$ the temperature evolution of a region of the same \ac{OUT} without defects is subtracted. As a second post-processing method, the amplitude and phase data for a \acf{PPT} of $T_\text{diff}$ for a evaluation frequency of $f_\text{PPT}=\SI{0.1}{\hertz}$ (found to feature the highest contrast for this dataset) is shown\cite{IbarraCastanedo2004}.\\

\begin{figure*}
\centering
\includegraphics[scale=1]{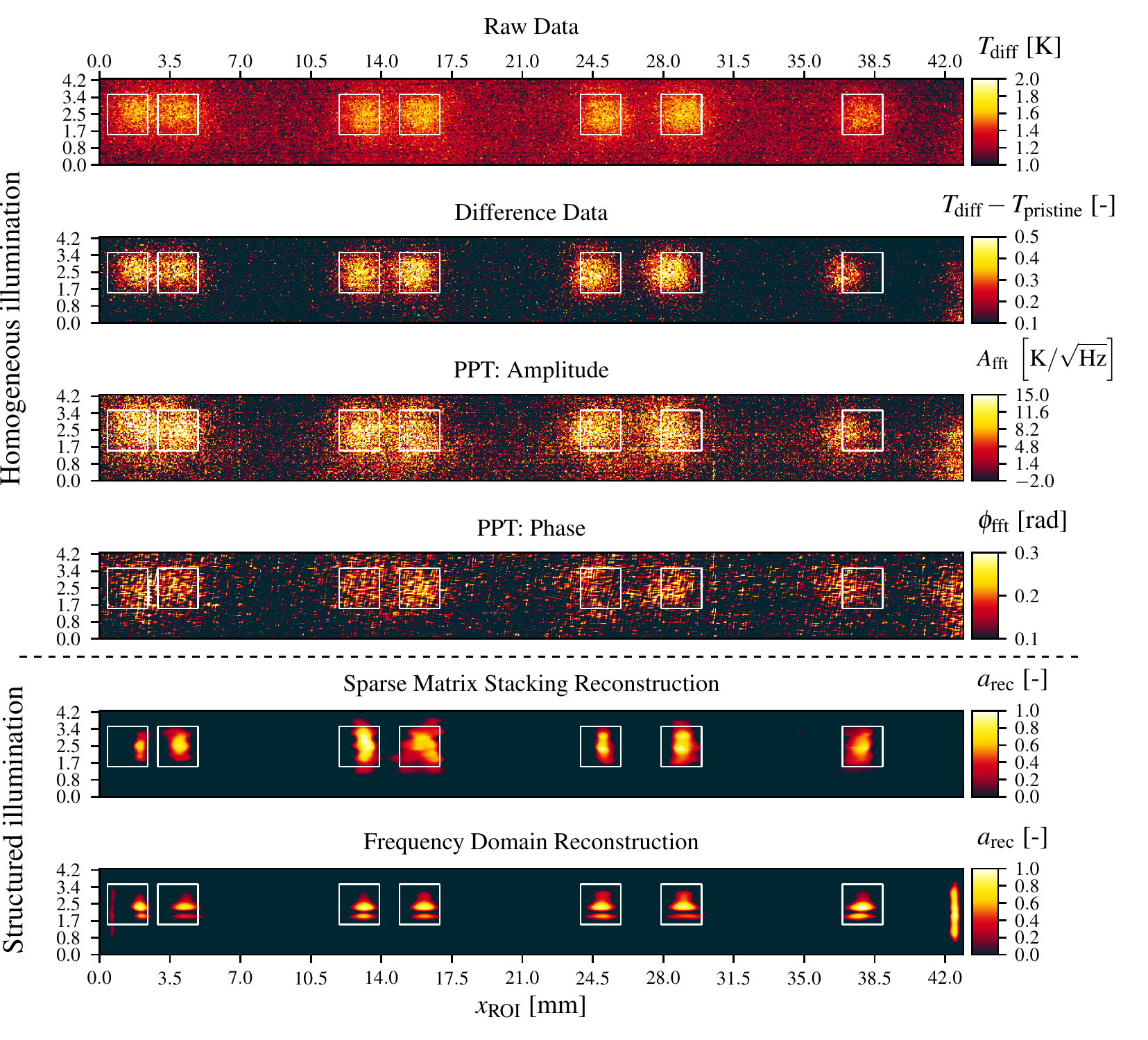}
\caption{Defect detection performance comparison of the proposed \ac{PSR} reconstruction methods with conventional methods. From top to bottom: Unprocessed measured data $T_\text{diff}$ from \reffigl{fig:res_flash} for the chosen \ac{ROI}; Difference thermogram showing the difference of the measured data and the measurement data for a different defect-free \ac{ROI}; Frequency amplitude data for a frequency of $f_\text{PPT}=\SI{0.1}{\hertz}$ for $T_\text{diff}$ processed by \ac{PPT}; Frequency phase data for a frequency of $f_\text{PPT}=\SI{0.1}{\hertz}$ for $T_\text{diff}$ processed by \ac{PPT}; $a_\text{rec}$ as reconstructed by \ac{PSR} reconstruction with the proposed sparse matrix stacking method; $a_\text{rec}$ as reconstructed by \ac{PSR} reconstruction in the frequency domain. The white boxes in each plot indicate the true defect positions.}
\label{fig:res_comp}
\end{figure*}

Since quantitatively comparing the increase in resolution capability comparing the sparse reconstruction results gained from \ac{PSR} reconstruction and dense measurement data acquired from conventional methods is not trivially possible, we limit ourselves to a qualitative comparison only where our main criterion comprises the separability of closely spaced defects by the method. As can be seen in \reffigl{fig:res_comp}, in the difference thermogram, as well as a \ac{PPT} analysis of $T_\text{diff}$ all defects are detected, but the smallest gap between defects ($d_\text{gap}=\SI{0.5}{\milli\metre}$) cannot be clearly identified with the available \ac{SNR} of the measurement. In this direct comparison the advantage of the sparse nature of the reconstruction of the \ac{PSR} reconstruction results can be clearly identified. Here all defects can be clearly separated. 

\section{Conclusion and Outlook}
Within this work, an experimental approach for expanding the capabilities of laser-based \ac{PSR} reconstruction to the detection of two-dimensional defects has been successfully introduced. Two numerical inversion methods for the reconstruction of the internal defect/inhomogeneity structure of an \ac{OUT} have been proposed and their reconstruction performance has been tested experimentally. Both proposed numerical methods have proven to make it possible to detect even closely positioned defects individually and both methods have provided an approximation of the defect shape. While the proposed sparse matrix stacking inversion method led to qualitatively better shape reconstruction, it took almost $50$ times more time to compute on modern computer hardware, making it less suited for the reconstruction of large \ac{ROI}s. The reconstruction in the frequency domain has proven to be more sensitive to the choice of regularization parameters and to imperfections in the measurements but takes less time to compute and is more memory efficient. Both methods are still equally affected by the lack of a suitable method for finding the best suited set weights for the regularizing terms programmatically in a feasible amount of time. Solving this issue and the improvement of the reconstruction quality of the defect shape is still part of ongoing research\cite{Ahmadi2020c}.\\

As the presented experimental approach can be recreated using pre-existing standard equipment a simple adaptation for existing industrial test processes is possible. Its capability of achieving \ac{SR} further allows for the detection of defects beyond the spatial resolution limit of the conventional methods presented. In addition, the sparse nature of the reconstruction results highly contrasts defective regions and allows for easy automation of the test result evaluation process. However, the experimental prerequisite to perform a large amount of sequential independent measurements with single laser-spot excitation to cover larger \ac{ROI}s is still to be improved for this method. While for testing scenarios demanding high-defect detection resolution at larger depths the increased testing effort can be justified, e.g., for high-performance parts for aerospace applications or for medical diagnostics, for high-volume/low-cost testing scenarios the experimental complexity still needs to be improved. Here the projection of two-dimensional pixel patterns using latest laser-projector technology shows promising results to significantly reduce the necessary number of measurements\cite{Lecompagnon2021b}.

\FloatBarrier

\begin{acknowledgments}
None.
\end{acknowledgments}

\section*{Conflict of Interest}
The authors have no conflicts to disclose.

\section*{Data Availability Statement}
The data that support the findings of this study are available from the corresponding author upon reasonable request.\\

\bibliography{refs}

\end{document}